\documentclass[print, sort&compress]{elsarticle}

\usepackage{lineno}

\usepackage{graphicx}
\usepackage{dcolumn}
\usepackage{bm}
\usepackage{upgreek}
\usepackage{hyperref}
\usepackage{amsmath}
\usepackage{amssymb}
\usepackage{multirow}
\usepackage[margin=1.3in]{geometry}
\let\oldequation\equation
\let\oldendequation\endequation

\renewenvironment{equation}
  {\linenomathNonumbers\oldequation}
  {\oldendequation\endlinenomath}

\usepackage{multicol} 

\usepackage{nomencl} 

\makenomenclature

\usepackage{hyperref}
\usepackage{soul,color,xcolor}
\usepackage{etoolbox}
\renewcommand\nomgroup[1]{%
  \item[\bfseries
  \ifstrequal{#1}{G}{Greek symbols}{%
  \ifstrequal{#1}{S}{Subscripts}{}}%
]}

\newcommand{\We}{W\!e}%
\newcommand{\Oh}{O\!h}%
\newcommand{\De}{D\!e}%
\renewcommand{\Oh}{\text{Oh}}%
\renewcommand{\We}{\text{We}}%
\renewcommand{\Re}{\text{Re}}%
\renewcommand{\De}{\text{De}}%
\journal{Journal of Colloid and Interface Science}

\begin{document}

\begin{frontmatter}

\title{Singular jets during droplet impact on superhydrophobic surfaces}

\author[SKLE]{Xiaoyun Peng}
\author[SKLE,PES]{Tianyou Wang}
\author[SKLE]{Feifei Jia}
\author[SKLE]{Kai Sun}
\author[SCHE]{Zhe Li}
\author[SKLE,PES]{Zhizhao Che\corref{cor1}}
\cortext[cor1]{Corresponding author.
}
\ead{chezhizhao@tju.edu.cn}
\address[SKLE]{State Key Laboratory of Engines, Tianjin University, Tianjin, 300350, China.}
\address[PES]{National Industry-Education Platform of Energy Storage, Tianjin University, Tianjin, 300350, China}
\address[SCHE]{State Key Laboratory of Chemical Engineering, Tianjin University, Tianjin, 300350, China.}

\begin{abstract}
\emph{Hypothesis}: The impact of droplets is prevalent in numerous applications, and jetting during droplet impact is a critical process controlling the dispersal and transport of liquid. New jetting dynamics are expected in different conditions of droplet impact on super-hydrophobic surfaces, such as new jetting phenomena, mechanisms, and regimes.\\
\emph{Experiments}: In this experimental study of droplet impact on super-hydrophobic surfaces, the Weber number and the Ohnesorge number are varied in a wide range, and the impact process is analyzed theoretically.\\
\emph{Findings}: We identify a new type of singular jets, i.e., singular jets induced by horizontal inertia (HI singular jets), besides the previously studied singular jets induced by capillary deformation (CD singular jets). For CD singular jets, the formation of the cavity is due to the propagation of capillary waves on the droplet surface; while for HI singular jets, the cavity formation is due to the large horizontal inertia of the toroidal edge during the retraction of the droplet after the maximum spreading. Key steps of the impact process are analyzed quantitatively, including the spreading of the droplet, the formation and the collapse of the spire, the formation and retraction of the cavity, and finally the formation of singular jets. A regime map for the formation of singular jets is obtained, and scaling relationships for the transition conditions between different regimes are analyzed.
\end{abstract}

\begin{keyword}
\texttt {
Droplet impact \sep
Singular jet \sep
Cavity retraction \sep
Superhydrophobic surfaces
}
\end{keyword}

\end{frontmatter}


\def \scaleSize {0.8}
\def \scaleSiz2 {0.6}
\section{Introduction}
Droplet impact is a ubiquitous process in nature and is also an important problem in many industrial applications, such as pesticide spraying \cite{Yarin2006}, inkjet printing \cite{Basaran2013, Holman2002, Park2006}, and spray cooling \cite{Josserand2016, Lin1998DropImpact, Zhang2014}. Depending on the impact conditions, there are several outcomes after the impact, including deposition, bouncing, splashing, and jetting \cite{Abolghasemibizaki2018, Ahmed2018, Ashoke2017, Cervantes2020, Holman2002, Hu2021, Motzkus2011, Sikalo2005, Ven2023, Yarin2006}. Among them, jetting during droplet impact is a critical process in controlling droplet behavior in the aforementioned applications. Because of the huge transient inertia and flow focusing near the free surface \cite{Lai2018, Lin1998DropImpact, Michon2017, Siddique2020}, jetting is critical for the dispersal and transport of liquid \cite{Lohse2022} in droplet impact processes.

Jets can be produced after the impact of droplets on liquid or solid surfaces, as well as in many other processes, such as bubble bursting at a free liquid surface \cite{Ganan2017, Gordillo2019, Lai2018, Lee2021}, pinch-off of a droplet or a bubble from an underwater nozzle \cite{Gekle2010, Gordillo2010}, or interface oscillation \cite{Benjamin2000, Elise2002}. For droplet impact processes \cite{Scheller1995, Chen2018, Clanet2004, Chijioke2005, Lin2018, Nguyen2020, Pack2019}, jets are often formed during droplet impact on a liquid pool \cite{Deng2007, Lee2021, Michon2017, Pan2010, Hasan1990, Thoroddsen2018, Yang2020} or a solid substrate \cite{Bartolo2006, Zhang2022, Josserand2016, Lin2018, Nguyen2020, Pasandideh1996, Siddique2020, Yamamoto2018}. For droplet impact on a liquid pool \cite{Michon2017, Thoroddsen2018, Yang2020}, a crater forms by the initial inertia. Then, the crater retracts by surface tension force \cite{Yarin2006}, or suddenly collapses by capillary waves convergence or by pure inertia. When the crater retracts by surface tension force, a thick jet forms from the center of the crater, which is called a Worthington jet \cite{Gekle2010, Gordillo2010, Yamamoto2018, Yarin2006}. In contrast, when the crater collapses under capillary force or pure inertia by capillary wave convergence, a singularity (due to a special morphology like a collapsing depression) of the crater bottom (at which the surface curvature diverges due to the capillary waves converge and collapse) induces high kinetic energy of fluid along the central axis, forming a thin high-speed jet forms with a series of satellite droplets formation, which is called a singular jet \cite{Bartolo2006, Benjamin2000, Brenner2000, Chen2017, Thoroddsen2018, Yang2020}. For droplet impact on a solid substrate \cite{Siddique2020, Yamamoto2018}, not only the initial inertia, but also the substrate affects the cavity formation. The substrate limits the development of the cavity in the vertical direction and induces a liquid film \cite{Renardy2003, Roisman2009} in the center. When the solid substrate is hydrophilic, the cavity retracts by the rim retraction, and the liquid flows into the center, producing a Worthington jet \cite{Pasandideh1996, Siddique2020, Yamamoto2018}. In contrast, when the solid substrate is super-hydrophobic, the low surface energy of the substrate induces the violent deformation of the interface and the formation of a thin liquid film. Therefore, the collapse of the cavity pushes the thin liquid film forming a singular jet \cite{Bartolo2006, Benjamin2000, Brenner2000, Chen2017, Thoroddsen2018, Yang2020}.

The interface deformation for singular jets formation during the impact of droplets on liquid pools or super-hydrophobic surfaces is a complex process \cite{Siddique2020, Thoroddsen2018, Yamamoto2018, Yang2020}. Singular jets are produced from singularities \cite{Benjamin2000, Brenner2000} in the crater/cavity collapse step during droplet impact on a liquid pool \cite{Thoroddsen2018, Yang2020} or a super-hydrophobic surface \cite{Bartolo2006, Nguyen2020}. For droplet impact on a liquid pool, the complexity of singular jets depends on the interface morphology in the retraction stage \cite{Thoroddsen2018, Yang2020}. When the droplet and the liquid pool are miscible \cite{Thoroddsen2018} or immiscible \cite{Yang2020}, the shape of the crater bottom is like a dimple \cite{Thoroddsen2018, Yang2020}, and the parameter space involved is complex. The dimple shape depends on whether the process is in single collapse regimes (i.e., capillary-inertial regime and pure inertial regime) or the cross-over collapse regime \cite{Thoroddsen2018, Yang2020}. The dimple shape determines the singular jet velocity, and the fastest jet forms with a toroidal micro-bubble entrainment from the bottom of the dimple \cite{Thoroddsen2018, Yang2020}. Therefore, as the characteristic phenomenon in singular jet formation, the singular jet formation with bubble entrainment has also been studied \cite{Deng2007, Hasan1990, Thoroddsen2018}, which shows two boundaries. One boundary is from bubble entrainment formation, which is determined by the comparison of horizontal and longitudinal characteristic length scales. The horizontal characteristic length scale depends on the balance of inertia and surface tension force, and the longitudinal characteristic length scale depends on the balance of inertia and gravity \cite{Deng2007, Hasan1990, Thoroddsen2018}. The other boundary is for the critical condition of bubble entrainment disappearance, and a curvature reversal disappearance at the bottom is determined by the comparison between the time of capillary wave at the bottom of the crater and the time of reaching the maximum crater scale \cite{Deng2007, Hasan1990, Thoroddsen2018}.

For droplet impact on the super-hydrophobic surface, the solid substrate limits the formation and collapse of the cavity \cite{Bartolo2006, Zhang2022}. In the retraction stage, a bubble could be trapped under the severe capillary deformation of the cavity \cite{Bartolo2006, Zhang2022, Lin2018, Nguyen2020, Guo2020, Chen2017}. The physical mechanism of the cavity collapse can be explained by the Rayleigh-Plesset equation \cite{Bartolo2006, Chen2017, Milton1977}, which describes the cylindrical free surface flows in pinch-off dynamics. According to the Rayleigh-Plesset equation, a 1/2-power law for the evolution of the cavity radius has been obtained \cite{Bartolo2006, Chen2017}. The relationship between the velocity and the radius of the singular jet can also be obtained based on mass and kinetic energy fluxes \cite{Bartolo2006, Chen2017}. However, many aspects of singular jet formation are still unclear, including the mechanisms and the conditions of singular jet formation.

In this study of droplet impact on super-hydrophobic surfaces, we find a new type of singular jets (i.e., singular jets induced by horizontal inertia), besides the previously studied singular jets, which are named singular jets induced by capillary deformation in this study. By experimental measurement and theoretical analysis, the mechanisms of the formation of the singular jets are unveiled. To explore the details of the two types of singular jets, the key features during the spreading and retraction of the droplet are considered. The transition conditions for the occurrence of the singular jets are analyzed.

\section{Materials and methods}\label{sec:2}
\begin{figure}
  \centering
  \includegraphics[width=\columnwidth]{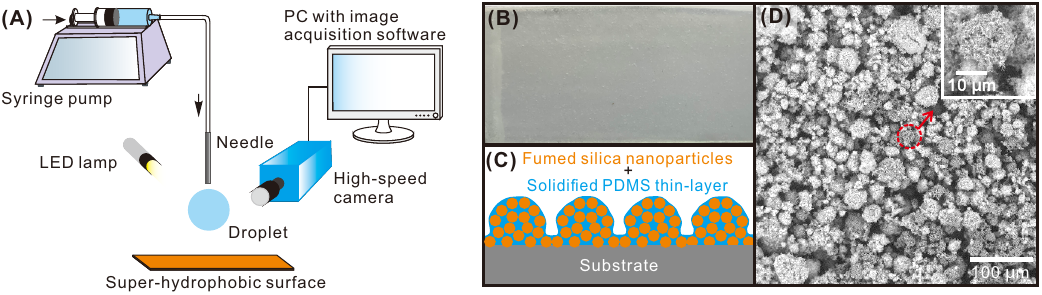}
  \caption{(A) Schematic diagram of the experimental setup for the impact of droplets on the super-hydrophobic surface. (B) Optical photos of the super-hydrophobic surface. (C) Schematic diagram of the super-hydrophobic surface. (D) SEM images of the super-hydrophobic surface, inset: a magnification of the surface.}\label{fig:01}
\end{figure}

The experimental setup is sketched in Figure \ref{fig:01}(A). Droplets were slowly released from a needle pushed by a syringe pump (Harvard Apparatus, Pump 11 elite Pico plus) and subsequently free fell on a super-hydrophobic surface. Different impact velocities of the droplet were obtained by adjusting the droplet's falling height. A high-speed camera (Phantom V1612, America) was used to record the impact process at $5,400$ -- $100,000$ frames per second (fps), and with the spatial resolution ${24}-{47}$ $\upmu\text{m}/\text{pixel}$. To obtain high-quality images, we used a 100 mm lens (Nikon AF 100 mm f/2.8D) with a small aperture (F11) on the high-speed camera. A high-power LED lamp was used to illuminate the process for high-speed imaging. To obtain quantitative information on the impact process, the high-speed images were analyzed via a customized image processing program. Regarding the uncertainty of the measurement of the droplet diameter, it is mainly due to the spatial resolution of high-speed imaging. In the experiment, the uncertainty is $\pm{0.8}\%-\pm{2}\%$ for the droplet diameter, $\pm{0.3}\%-\pm{0.6}\%$ for the droplet velocity, $\pm{7.2}\%-\pm{50}\%$ for the singular jet radius (the singular jet radius is very small compared with the droplet diameter), and $\pm{4.5}\%-\pm{14.3}\%$ for the singular jet velocity. In addition, the uncertainties of the We and Oh numbers are $\pm{1}\%-\pm{2.3}\%$ and $\pm{0.4}\%-\pm{1}\%$, respectively.

The super-hydrophobic surfaces were prepared according to the procedure reported in Ref. \cite{Lee2019}, as shown in Figure \ref{fig:01}(A). Briefly, the super-hydrophobic surface was prepared by dip-coating on a flat glass slide with fumed silica nanoparticles (AEROSIL R972 supplied by DEGUSSA) and PDMS (Sylgard 184). The PDMS was firstly diluted with hexane (10\% [w/v]) to improve the fluidity, which can easily dip-coat on a smooth glass slide. After the evaporation of most hexane at room temperature, a PDMS adhesive thin-layer with a thickness of ${3.5}\pm {0.7}$ $\upmu\text{m}$ was deposited. Then, dry nanoscale powders (AEROSIL R972) were directly sprinkled onto the PDMS adhesive thin layer through a sieve. The excess powder on the surface was removed and recycled by an aspirator. Scanning electron microscopy (SEM) images of the super-hydrophobic surface are shown in Figure \ref{fig:01}(D). The surface has microscale aggregates of ${25.6}\pm {7.4}$ $\upmu\text{m}$ in diameter, which is the raised structure composed of fumed silica nanoparticles (${16}$ $\text{nm}$ in diameter, orange particles in Figure \ref{fig:01}(C)) and PDMS adhesive thin layer (blue thin layer in Figure \ref{fig:01}(C)). The contact angle of the super-hydrophobic surface is ${154}$${^\circ}$. The sliding angle (i.e., the difference between advancing and receding contact angles, which indicate contact angle hysteresis) of the superhydrophobic surface is ${0.4}\pm {0.1}$${^\circ}$. The surface roughness of the superhydrophobic surface is ${14.3}\pm {0.2}$ $\upmu\text{m}$, which was measured via a laser confocal microscope \cite{Lee2019}. To verify that the occurrence of the singular jets depends on the droplet impact dynamics rather than the particularity of superhydrophobic surfaces, we performed additional experiments using different types of superhydrophobic surfaces. The results show that the singular jet formation is unaffected by the particularity of superhydrophobic surfaces (see Section S2 in Supplementary Material for details).

Droplet impact experiments were carried out with DI water and glycerol-water mixtures at room temperature, and the fluid properties are provided in Table S1 in Supplementary Material. By varying the concentration of the glycerol-water solutions, we can consider the effect of the fluid viscosity while keeping the surface tension and the density almost constant.

The impact dynamics are mainly controlled by inertial forces, viscous forces, and capillary forces. Therefore, we select the Weber number ${\We}={{\rho} {{D}_{0}}{{U}_{0}^{2}}}/{\sigma }$ and the Ohnesorge number ${\Oh}={\mu }/{\sqrt{{\rho} {{D}_{0}}{\sigma} }}$ to characterize the impact process, where ${{U}_{0}}$ and ${{D}_{0}}$ are the impact velocity and the droplet diameter; ${\rho} $, ${\sigma}$, and ${\mu }$ are the density, surface tension, and dynamic viscosity of the liquid, respectively. The Weber number ${\We}$ represents the ratio between the inertia and the surface tension force, and the Ohnesorge number ${\Oh}$ represents the relative importance of the viscous force. They are varied in a wide range in this study (i.e., ${\We}$ from ${2.68}$ to ${97.4}$ and ${\Oh}$ from ${2.03}\times {{10}^{-3}}$ to ${4.14}\times {{10}^{-2}}$).

\section{Results and discussion}\label{sec:3}
\subsection{Two types of singular jets}\label{sec:31}

\begin{figure}
  \centering
  \includegraphics[width=0.8\columnwidth]{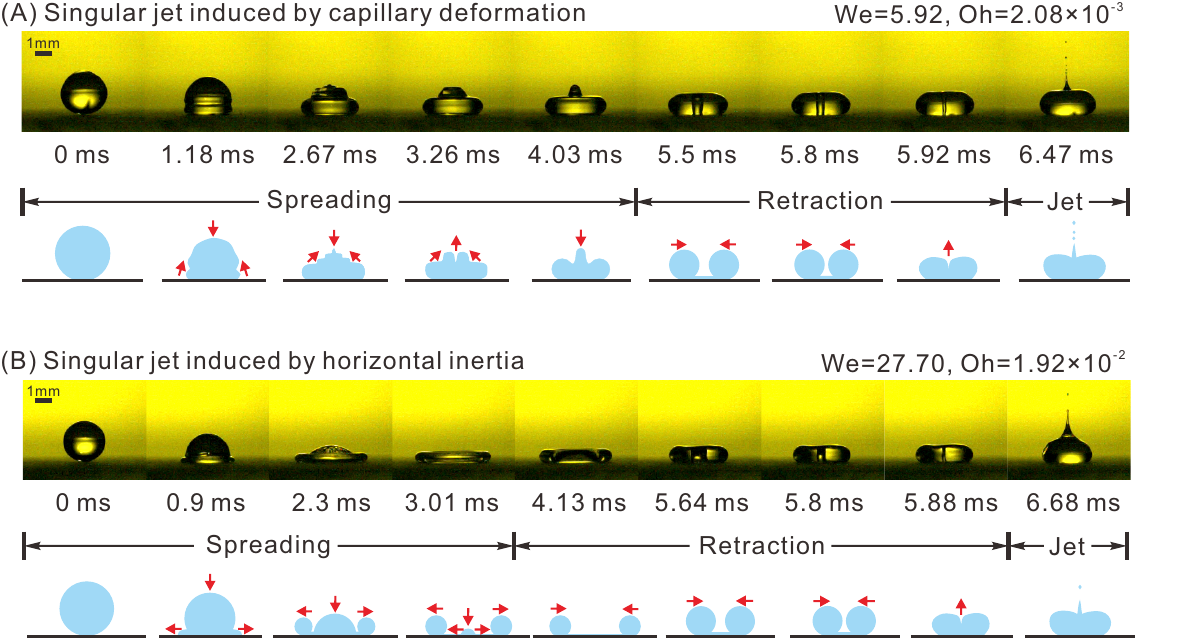}
  \caption{Two types of singular jets during droplets impact on the super-hydrophobic surface. (A) Singular jet induced by capillary deformation (i.e., CD singular jet). The corresponding experimental setting is ${\We}={5.92}$, ${\Oh}={2.08}\times  {10}^{ -3}$. (B) Singular jet induced by horizontal inertia (i.e., HI singular jet). The corresponding experimental setting is ${\We}={27.70}$, ${\Oh}={1.92}\times {10}^{-2}$. In each panel, the first row is experimental images and the second row is the schematic drawing. Video clips of these processes are available as Supplementary Material as Movies 1 and 2.}\label{fig:02}
\end{figure}

From the experiment images, we find two types of singular jets during the droplet impact on the super-hydrophobic surface, as shown in Figure \ref{fig:02}. The analysis of the mechanisms of singular jet production shows that the two types of singular jets are induced by capillary deformation (CD singular jet) and by horizontal inertia (HI singular jet), respectively, which will be discussed separately.

The first type of the singular jet is formed due to capillary deformation, as shown in Figure \ref{fig:02}(A). When the droplet contacts the substrate, capillary waves propagate from the droplet/substrate interface, and droplet spreading is remarkably affected by the capillary waves. The droplet deforms as the capillary waves propagate upward to the top, forming the shape of a pyramid. Due to the large surface tension force (i.e., large curvature) at the tip of the pyramid, a small cavity can be produced at the top of the droplet. At this point, the capillary waves continuously converge to the center of the droplet, and the small cavity (i.e., large curvature) retracts under the large surface tension force. Afterward, a spire is formed on the droplet. Because of the surface tension force induced by the spire, the spire collapses downward, forming a cavity. Due to the surface singularity (a collapsing depression \cite{Benjamin2000}) caused by the cavity and the thin film, the cavity bottom has a surface curvature divergence. Due to the divergence of the surface curvature (suddenly increasing curvature) of the cavity and the corresponding large surface tension force, the cavity retracts inwards, and a singular jet is finally formed when the cavity collapses at the center.

In addition to singular jets induced by capillary deformation, we identify another type of singular jet, which is induced by horizontal inertia, as shown in Figure \ref{fig:02}(B). After the impact, there is no strong capillary wave on the surface of the droplet during droplet spreading, which is due to the inhibition effect by the viscous dissipation on the capillary wave propagation. Therefore, after the droplet contacts the substrate, the droplet first deforms to the shape of a cap under inertia. The top surface of the droplet continues to move download until the formation of a liquid film at the center. In this way, the thin central film and the toroidal edge form a cavity (a collapsing depression \cite{Benjamin2000}, where have a surface singularity at the bottom). After that, the toroidal edge retracts under the surface tension force and transfers the surface energy into the horizontal inertia. Then, large horizontal inertia from the droplet edge pushes the cavity to collapse toward the axis. The cavity recoils under the large horizontal inertia and surface singularity of the cavity, eventually forming a singular jet.

A major difference between the two types of singular jets is the mechanism of the formation of the cavity, which retracts to produce the singular jet. For the CD singular jet, the cavity formation is due to the propagation of the capillary wave on the surface of the droplet, which induces strong surface vibration at the apex of the droplet. In contrast, for the HI singular jet, the cavity formation is by the thin central film and the toroidal edge, which has large horizontal inertia during the retraction of the droplet after the maximum spreading. The singular jets reported in many studies \cite{Bartolo2006, Chen2017, Chen2018, Lin2021, Lin2018} belong to the first type (i.e., CD singular jet). However, to the best of our knowledge, the HI singular jet has never been reported in the literature.

In the next two sections (Sections \ref{sec:32} and \ref{sec:33}), we will analyze the impact processes by considering key features of the spreading and retraction stages, such as the maximum spreading diameter, the maximum cavity diameter, the center film thickness, and the collapse of the cavity. The results of the analysis will be used to determine the transition conditions between different regimes in Section \ref{sec:34}.

\subsection{Spreading stage}\label{sec:32}
Spreading is the first stage of the impact process. When a droplet contacts the super-hydrophobic surface, the flow in the droplet will develop horizontally under inertia, and the droplet is deformed by the surface tension force during the spreading. As the initial kinetic energy of the droplet is converted into surface energy, the droplet gradually reaches its maximum spreading diameter. In the final stage of droplet spreading, the center of the droplet further deforms and forms a cavity by the propagation of the capillary wave (for the scenario of the CD singular jet) or by the deformation of the thin central film and the toroidal edge (for the scenario of the HI singular jet). The main features in the spreading stage are discussed in this section.

\subsubsection{Maximum spreading diameter}\label{sec:321}
\begin{figure}
  \centering
  \includegraphics[width=\columnwidth]{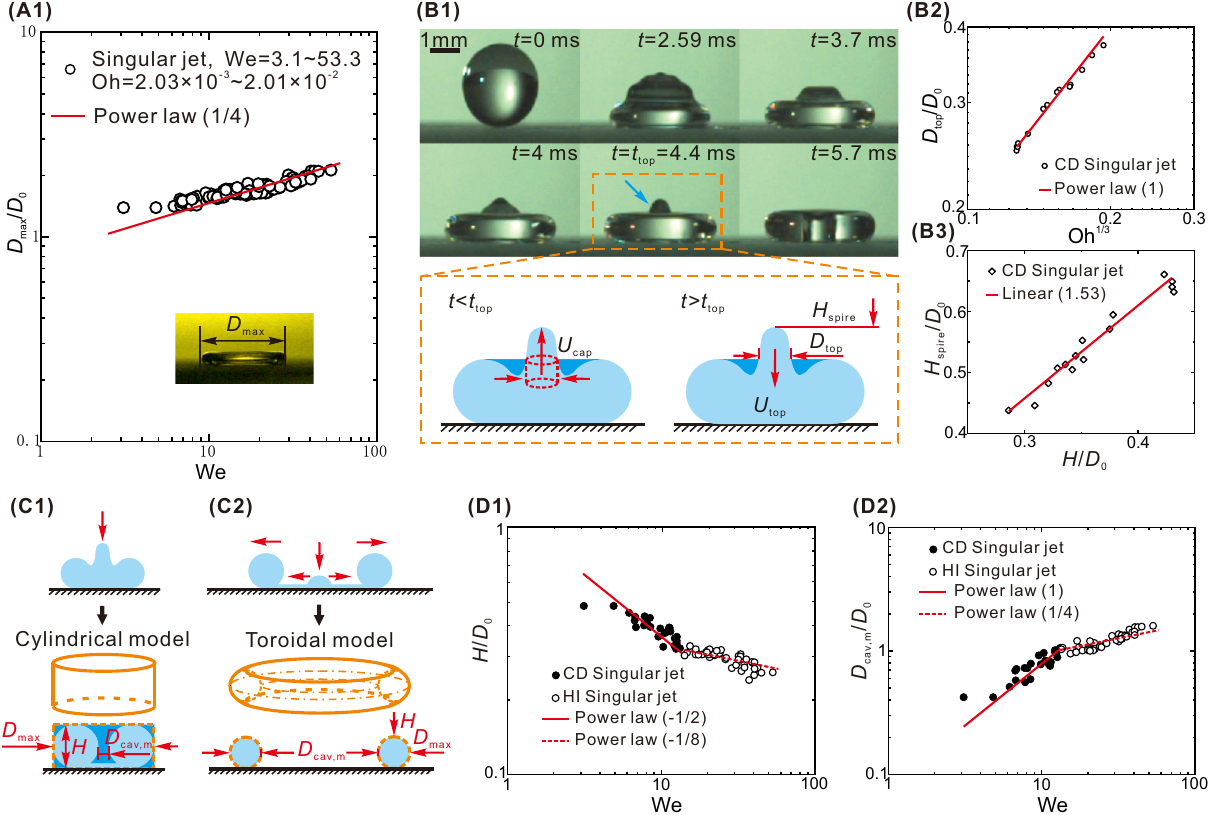}
  \caption{(A1) Dimensionless maximum spreading diameter as a function of the Weber number. The red solid line is ${{{D}_{\max }}}/{{{D}_{0}}}={0.82}{{\We}^{{1}/{4}}}$ (i.e., Eq. (\ref{eq:01})). (B1) Formation and collapse of the spire. A schematic diagram of the critical state before and after the spire formation is shown in the dashed box, where ${{t}_\text{top}}$ is the time for the formation of a center liquid column; ${{U}_\text{cap}}$ is the velocity of the capillary wave; ${{U}_\text{top}}$ is the velocity of the top surface of the spire; ${{D}_\text{top}}$ is the top diameter of the spire; ${{H}_{\text{spire}}}$ is the height of the spire. (B2) Dimensionless top diameter of the spire as a function of the Ohnesorge number. The solid line is ${{D}_\text{top}}/{{D}_{0}}={2.0}{\Oh}^{{1}/{3}}$ (i.e., Eq. (\ref{eq:05})). (B3) Dimensionless height of the spire as a function of the dimensionless droplet thickness ${H}/{{{D}_{0}}}$. The solid line is ${{{H}_{\text{spire}}}}/{{{D}_{0}}}={1.53}{H}/{{{D}_{0}}}$ (i.e., Eq. (\ref{eq:07})). (C1-C2) Two geometrical models to estimate the droplet volume. (C1) The cylindrical model: the cavity is small and the droplet volume can be estimated as a cylinder. (C2) The toroidal model: the center liquid film is thin and the droplet volume can be estimated as a toroidal. (D1) Dimensionless droplet thickness ${H}/{{{D}_{0}}}$ as a function of Weber number. The red solid line is ${H}/{{{D}_{0}}}={1.14}{\We}^{-1/2}$ (i.e., Eqs.\ (\ref{eq:12})), and the red dash line is ${H}/{{{D}_{0}}}={0.44}{{\We}^{{-1}/{8}}}$ (i.e., Eq. (\ref{eq:18})), in which the coefficients are fitted from experimental data based on Eqs.\ (\ref{eq:11}) and (\ref{eq:17}), respectively. (D2) Dimensionless maximum cavity diameter ${{{D}_\text{cav,m}}}/{{{D}_{0}}}$ as a function of Weber number. The red solid line is ${{{D}_\text{cav,m}}}/{{{D}_{0}}}={0.079}{\We}$ (i.e., Eq. (\ref{eq:15})), and the red dash line is ${{{D}_\text{cav,m}}}/{{{D}_{0}}}={0.53}{\We}^{{1}/{4}}$ (i.e., Eqs. (\ref{eq:20})), in which the coefficients are fitted from experimental data based on Eqs.\ (\ref{eq:14}) and (\ref{eq:19}), respectively.
}\label{fig:03}
\end{figure}

The maximum spreading diameter ${{D}_{\max }}$ (see Figure \ref{fig:03}(A1)) is an important parameter to characterize the droplet deformation after the droplet impact on a solid substrate. Many models \cite{Scheller1995, Clanet2004, Eggers2010, Josserand2016, Chijioke2005, Pasandideh1996, Roisman2009} have been proposed to consider the maximum spreading diameter, and the theoretical model by Clanet et al.\ \cite{Clanet2004} is adopted in this study, which shows good quantitative agreement with experimental data for hydrophobic substrates and can unify most existing models \cite{Scheller1995, Clanet2004, Eggers2010, Josserand2016, Chijioke2005, Pasandideh1996, Roisman2009}. An impact number (i.e., ${P}={{{{\We}}/{{\Re}}}^{{4}/{5}}}$) is defined in the capillary and viscous regimes (i.e., ${{{D}_{\max }}}/{{D}_{0}}\sim {\We}^{{1}/{4}}$ for the capillary regime, and ${{{D}_{\max }}}/{{D}_{0}}\sim {\Re}^{{1}/{5}}$ for the viscous regime) according to the conservation of mass and energy \cite{Clanet2004}. Through experimental data, the critical value of the impact number was found to be about ${P}={1}$, which can be used to distinguish the transition between the capillary and viscous regimes \cite{Clanet2004}. For the cases with singular jets in this study, the maximum impact number is ${0.46}$, indicating that the two types of singular jets both occur in the capillary regime of droplet spreading. Hence, the scaling for the capillary regime, ${{{D}_{\max }}}/{{D}_{0}}\sim {\We}^{1/4}$, can be used to describe the maximum spreading diameter. By fitting with our experimental data, we can obtain
\begin{equation}\label{eq:01}
  {{{D}_{\max }}}/{{{D}_{0}}}={0.82}{{\We}^{{1}/{4}}}.
\end{equation}
As shown in Figure \ref{fig:03}(A1), the scaling agrees well with our experimental data.

\subsubsection{Spire formation under capillary waves}\label{sec:322}
For the formation of a CD singular jet, a spire is formed at the apex of the droplet (denoted by the blue arrow at ${4.4}$ $\text{ms}$ in figure \ref{fig:03}(B1)) as the capillary wave propagates to the apex of the droplet, forming a center liquid column \cite{Chen2017, Lin2021}. Then, the spire collapses downwards suddenly, producing a cavity (${5.7}$ $\text{ms}$ in Figure \ref{fig:03}(B1)). The downward velocity of the top surface at the spire ${{U}_\text{top}}$ increases suddenly at the moment of collapse, and becomes larger than the initial velocity of the top surface (i.e., the initial top velocity is at the same magnitude as the impact velocity ${{U}_{{0}}}$) \cite{Pack2019}.

To analyze the reason for the sudden collapse of a spire, we first consider the mechanism of the spire formation. The spire formation is similar to the liquid jet formation during bubble bursting \cite{Gordillo2019, Lai2018, Lee2021}, where a spire is formed by the capillary wave converging. For the liquid jet formation during bubble bursting, the radius of the jet ${R}$ and the jet speed ${U}$ scale as ${R}/{{{l}_{\mu }}}\sim {{({U}/{{{U}_{\mu }}})}^{-{5}/{3}}}$ based on momentum conservation \cite{Ganan2017}, where ${{l}_{\mu }}\sim {{{\mu }^{2}}}/{{\rho} {\sigma} }$ and ${{U}_{\mu }}\sim {\sigma }/{\mu }$. For the center liquid column in this study, we can consider the momentum conservation of the liquid column similarly. The spire diameter and its velocity follow
\begin{equation}\label{eq:02}
  {{{R}_\text{top}}}/{{{l}_{\mu }} }\sim {{( {{{U}_\text{cap}}}/{{{U}_{\mu }}})}^{-{5}/{3}}},
\end{equation}
where ${{R}_\text{top}}$ is the spire radius, and ${{U}_\text{cap}}$ can be estimated from the velocity of the capillary wave
\begin{equation}\label{eq:03}
  {{U}_\text{cap}}\sim {{\left( {{2}\sigma }/{\rho {{D}_{0}}} \right)}^{{1}/{2}}}.
\end{equation}
By substituting Eq.\ (\ref{eq:03}) into Eq.\ (\ref{eq:02}), and then after rearrangement, we can get the spire diameter ${{D}_\text{top}}$
\begin{equation}\label{eq:04}
 {{{D}_\text{top}}}/{{{D}_{0}}}\sim {{2}^{{1}/{6}}}{{\Oh}^{{1}/{3}}}.
\end{equation}
Then we plot the experimental data of ${{D}_\text{top}}$ against ${\Oh}^{1/3}$ in Figure \ref{fig:03}(B2). By fitting the experimental data using the scaling of Eq.\ (\ref{eq:04}), we can obtain
\begin{equation}\label{eq:05}
  {{{D}_\text{top}}}/{{{D}_{0}}}={2.0}{{\Oh}^{{1}/{3}}}.
\end{equation}
The comparison between the experimental data with the scaling shows good agreement, which confirms that the spire is formed by the convergence of the capillary wave.

For the formation of the spire, it depends on the development of capillary waves on the surface of the droplet. Therefore, the height of the spire ${{H}_{\text{spire}}}$ should be comparable with the thickness of the droplets ${H}$ (see in Figure \ref{fig:03}(D1)) after the spire collapsing. A larger ${H}$ will allow the development of a higher spire. For simplicity, we can have
\begin{equation}\label{eq:06}
 {{{H}_{\text{spire}}}}/{{{D}_{0}}}\sim {H}/{{{D}_{0}}}.
\end{equation}
Then we plot the experimental data of ${{{H}_{\text{spire}}}}/{{{D}_{0}}}$ against ${H}/{{{D}_{0}}}$ in Figure \ref{fig:03}(B3). By fitting the experimental data using the scaling in Eq.\ (\ref{eq:06}), we can obtain
\begin{equation}\label{eq:07}
  {{{H}_{\text{spire}}}}/{{{D}_{0}}}={1.53}{H}/{{{D}_{0}}}.
\end{equation}

After the formation of the spire, two possible mechanisms could drive the downward motion of the spire. One mechanism is the continuation of the downward movement of the droplet liquid due to the initial inertia of the droplet impact. The other mechanism is the effect of the surface tension force created by the large curvature of the spire. If the top surface of the spire is dominated by inertia, it should not be affected by the capillary deformation near the substrate, and preserve the initial velocity of the droplet impact (${{U}_\text{top}}={{U}_{0}}$). From high-speed images, we can see that the downward speed of the spire ${{U}_\text{top}}$ is much larger than the initial velocity of the droplet impact ${{U}_{0}}$. Therefore, we can safely exclude this hypothesis.

In contrast, if the top surface velocity of the spire ${{U}_\text{top}}$ is produced by the surface tension force, we can estimate ${{U}_\text{top}}$ by considering the balance of terms in the momentum equation. Here, we assume that the viscous force can be neglected compared with the surface tension force and the inertial force due to the extremely small time ($\approx 1\ \text{ms}$) in the process. After the formation of the central liquid column ${{U}_\text{top}}$, the large local pressure at the spire produced by the large curvature drives the fluid downward acceleration at a characteristic velocity of ${{U}_\text{top}}$. By considering the momentum equation between the surface tension force term of the spire ${\sigma }/{{R}_\text{top}^{2}}$ and the inertial force term ${{\rho} {{U}_\text{top}^{2}}}/{{{H}_{\text{spire}}}}$, we have
\begin{equation}\label{eq:08}
  {\sigma }/{{R}_\text{top}^{2}}\sim {{\rho} {{U}_\text{top}^{2}}}/{{{H}_\text{spire}}}.
\end{equation}
By substituting Eqs.\ (\ref{eq:05}), (\ref{eq:07}), and (\ref{eq:12}) into Eq.\ (\ref{eq:08}), we can get
\begin{equation}\label{eq:09}
  {{U}_\text{top}}\sim {1.32}{{U}_{0}} {{\We}^{{-3}/{4}}} {{\Oh}^{{-1}/{3}}}.
\end{equation}

For the dimensionless numbers considered in this study (i.e., ${\We}={3.1}\sim {19.9}$, $\Oh=2.03\times {{10}^{-3}}\sim {8.99}\times {{10}^{-3}}$), the downward speed of the spire ${{U}_\text{top}}$ is greater than the initial velocity of the droplet impact (i.e., ${{U}_\text{top}}>{{U}_{0}}$). This is consistent with the experimental observation from the high-speed images. This can also confirm that the sudden collapse of the spire is due to the surface tension force, not the downward inertia of the droplet. The experimental data of ${{U}_\text{top}}$ cannot be obtained directly from the high-speed images since the downward movement of the spire is very fast. Eq.\ (\ref{eq:09}) will be used in the subsequent analysis which will be validated later.

\subsubsection{Maximum cavity diameter and the droplet thickness}\label{sec:323}
For droplet impact on superhydrophobic surfaces, the cavity formation will directly affect the retraction of the cavity and the formation of the singular jet. Therefore, it is necessary to study the mechanisms of cavity formation for the two types of singular jets. For the CD singular jet, the spire formed by the convergence of capillary waves will eventually collapse to form a cavity under the surface tension force (as analyzed in Section \ref{sec:322}). In contrast, for the HI singular jet, without the influence of capillary waves, the top surface of the droplet is not deformed and only moves downward to form a cavity under inertia.

According to the aforementioned two mechanisms of the cavity, we propose two geometric models of the droplet deformation (i.e., a cylindrical model and a toroidal model), as shown in Figure \ref{fig:03}(C1) and Figure \ref{fig:03}(C2), respectively. For the cavity formed after the spire collapses, the cavity is generated under the influence of capillary wave deformation. In this condition, the Weber number is small (i.e., ${\We}={3.1}\sim {19.9}$), and the maximum spreading diameter ${{D}_{\max }}$ has a ${1/4}$ scaling relationship with the Weber number (i.e., Eq.\ (\ref{eq:01}) as discussed in Section \ref{sec:321}). Therefore, the maximum spreading diameter ${{D}_{\max }}$ is also small in this condition. The cavity diameter ${{D}_\text{cav}({t})}$ reaches its maximum ${{D}_\text{cav,m}}$ quickly after the droplet reaches its maximum spreading. The maximum cavity diameter ${{D}_\text{cav,m}}$ is much smaller than the maximum spreading diameter ${{D}_{\max }}$, and the droplet volume at this moment can be approximated by a cylinder (hence we call it the cylinder model, as illustrated in Figure \ref{fig:03}(C1)).

In contrast, for the cavity formed by the thin central film and the toroidal edge under the inertia, the influence of capillary waves is weak. In this case, the Weber number is much larger (i.e., ${\We}={13.5}\sim {53.3}$) than the first case, and according to the scaling relationship of the maximum spreading diameter in Eq.\ (\ref{eq:01}), the maximum spreading diameter ${{D}_{\max }}$ is much larger than the first case. The maximum cavity diameter ${{D}_\text{cav,m}}$ is close to the maximum spreading diameter ${{D}_{\max }}$, and the central film thickness ${h}$ is much smaller than the maximum cavity diameter ${{D}_\text{cav,m}}$. Therefore, the droplet volume, in this case, can be approximated by a toroidal (hence we call it the toroidal model, as illustrated in Figure \ref{fig:03}(C2)).

Using the two geometric models, we can further deduce the droplet thickness ${H}$ (Figure \ref{fig:03}(C1 and C2)) and the maximum cavity diameter ${{D}_\text{cav,m}}$ quickly after the droplet reaches the maximum spreading diameter ${{D}_{\max }}$. For the singular jet induced by capillary deformation, we can use the cylindrical model. Through the volume conservation of the cylinder and the initial spherical droplet, we have
\begin{equation}\label{eq:10}
  {\frac{4}{3}}{\pi }{{R}_{0}^{3}}\sim {\pi }{{R}_{\max }^{2}}{H},
\end{equation}
where ${{R}_{0}}$ is the droplet radius, and ${{R}_{\max }}$ is the maximum spreading radius. By substituting Eq.\ (\ref{eq:01}) into Eq.\ (\ref{eq:10}), we can get a relationship between the droplet thickness and the Weber number
\begin{equation}\label{eq:11}
  {H}/{{{D}_{0}}}\sim {0.99}{{\We}^{-{1}/{2}}}.
\end{equation}
Then we plot the experimental data of ${H}/{{{D}_{0}}}$ against ${\We}$ in Figure \ref{fig:03}(D1). By fitting the experimental data using the scaling of Eq.\ (\ref{eq:11}), we can obtain
\begin{equation}\label{eq:12}
  {H}/{{{D}_{0}}}={1.14}{{\We}^{{-1}/{2}}}.
\end{equation}

Meanwhile, from the balance between the horizontal droplet inertia and the surface tension force of the cavity
\begin{equation}\label{eq:13}
 {\sigma }{{R}_\text{cav,m}}\sim {\rho }{{U}_{0}^{2}}{{R}_{0}^{2}},
\end{equation}
where ${{R}_\text{cav,m}}$ is the maximum cavity radius. The maximum cavity diameter can be obtained
\begin{equation}\label{eq:14}
  {{{D}_\text{cav,m}}}/{{{D}_{0}}}\sim {\frac{1}{2}}{\We}.
\end{equation}
Then we plot the experimental data of ${{{D}_\text{cav,m}}}/{{{D}_{0}}}$ against ${\We}$ in Figure \ref{fig:03}(D2). By fitting the experimental data using the scaling of Eq.\ (\ref{eq:14}), we can obtain
\begin{equation}\label{eq:15}
  {{{D}_\text{cav,m}}}/{{{D}_{0}}}={0.079}{\We}.
\end{equation}

In contrast, for the HI singular jet, we use the toroidal model. Through the volume conservation of the toroidal and the initial spherical droplet, we have
\begin{equation}\label{eq:16}
 {\frac{4}{3}}{\pi} {{R}_{0}^{3}}\sim {\pi }{({H}/{2})^{2}}({2}{\pi }{{R}_{\max }}).
\end{equation}
By substituting Eq.\ (\ref{eq:01}) into Eq.\ (\ref{eq:16}), we can get
\begin{equation}\label{eq:17}
{H}/{{{D}_{0}}}\sim {0.51}{{\We}^{-{1}/{8}}}.
\end{equation}
Then we plot the experimental data of ${H}/{{{D}_{0}}}$ against ${\We}$ in Figure \ref{fig:03}(D1). By fitting the experimental data using the scaling of Eq.\ (\ref{eq:17}), we can obtain
\begin{equation}\label{eq:18}
 {H}/{{{D}_{0}}}=0.44{{\We}^{{-1}/{8}}}.
\end{equation}
For the toroidal shape of the droplet, the maximum cavity diameter ${{D}_\text{cav,m}}$ is approximately the maximum spreading diameter ${{D}_{\max }}$. Hence, using Eq.\ (\ref{eq:01}), we can get the maximum cavity diameter
\begin{equation}\label{eq:19}
 {{{D}_\text{cav,m}}}/{{{D}_{0}}}\sim {{{D}_{\max }}}/{{{D}_{0}}}\sim {{\We}^{{1}/{4}}}.
\end{equation}
Then we plot the experimental data of ${{{D}_\text{cav,m}}}/{{{D}_{0}}}$ against ${\We}$ in Figure \ref{fig:03}(D2). By fitting the experimental data using the scaling of Eq.\ (\ref{eq:19}), we can obtain
\begin{equation}\label{eq:20}
  {{{D}_\text{cav,m}}}/{{{D}_{0}}}={0.53}{{\We}^{{1}/{4}}}.
\end{equation}

From the curves of the dimensionless droplet thickness ${H}/{{{D}_{0}}}$ and the dimensionless maximum cavity diameter ${{{D}_\text{cav,m}}}/{{{D}_{0}}}$ shown in Figure \ref{fig:03}(D1 and D2), we can see two regimes, which corresponds to CD singular jets and HI singular jets, respectively. By comparing the experimental data and the models, we can find the models agree reasonably well with the experimental data. As the Weber number increases, the regime changes from CD to HI singular jets. Meanwhile, the power-law index decreases from ${1}$ to ${1/4}$ for the maximum cavity diameter, and the power-law index increases from ${-1/2}$ to ${-1/8}$ for the droplet thickness. Hence, our result suggests that the dominant force of the cavity formation will decrease as the regime changes with the Weber number. The cavity formation is determined by the inertia and the surface tension force for CD singular jets at small Weber numbers, while the cavity formation only depends on the inertia for HI singular jets as the Weber number increases.

\subsubsection{Center film thickness }\label{sec:324}
The central film formed by the spire collapse is a unique morphology due to the substrate constraint, which has a direct effect on the subsequent retraction of the droplet and the formation of singular jets. Because the liquid film is very thin and its flow speed is very slow, we can use the lubrication approximation \cite{DeenBook} to analyze the theoretical film thickness ${{h}_\text{th}}$. The central film moves downward under capillary force, and lubrication pressure in the liquid film acts as the resistance.

For CD singular jets, the dominant force during the collapse of the spire is the surface tension force. Balancing the surface tension pressure gradient ${{{P}_\text{i}}}/{L}\sim {{\sigma} {{R}_{0}}}/({{{R}_\text{top}^{2}}{{h}_\text{th}}})$ with the lubrication pressure gradient ${{{P}_\text{l}}}/{L}\sim {\mu} {{{U}_\text{top}}{{R}_{0}}}/({{{h}_\text{th}^{2}}{{R}_\text{cav,m}}})$ (i.e., incompressible viscous drainage) in the central film, where ${L}$ is a horizontal length scale of the thin central film, we have
\begin{equation}\label{eq:21}
  {\frac{{\sigma} {{R}_{0}}}{{{R}_\text{top}^{2}}{{h}_\text{th}}}}\sim {\mu} {\frac{{{U}_\text{top}}{{R}_{0}}}{{{h}_\text{th}^{2}}{{R}_\text{cav,m}}}}.
\end{equation}
By substituting Eqs.\ (\ref{eq:05}), (\ref{eq:09}), (\ref{eq:15}) into Eq.\ (\ref{eq:21}), the thickness of the central film can be obtained
\begin{equation}\label{eq:22}
  {{h}_\text{th}}\sim {33.42}{{D}_{0}}{{\We}^{{-5}/{4}}}{{\Oh}^{{4}/{3}}}.
\end{equation}
In contrast, for HI singular jets, the dominant force on the top surface of the droplet is inertia. Balancing the inertia pressure gradient ${{{P}_\text{i}}}/{L}\sim {{\rho} {{U}_{0}^{2}}}/{{{h}_\text{th}}}$ with the lubrication pressure gradient ${{{P}_\text{l}}}/{L}\sim {\mu} {{{U}_\text{top}}{{R}_{0}}}/({{{h}_\text{th}^{2}}{{R}_\text{cav,m}}})$ in the central film, we have
\begin{equation}\label{eq:23}
  {\frac{\rho {{U}_{0}^{2}}}{{{h}_\text{th}}}}\sim {\mu} {\frac{{{U}_\text{top}}{{R}_{0}}}{{{h}_\text{th}^{2}}{{R}_\text{cav,m}}}}.
\end{equation}
By substituting Eq.\ (\ref{eq:20}) and ${{U}_\text{top}}\sim {{U}_{0}}$ into Eq.\ (\ref{eq:23}), the thickness of the central film can be obtained
\begin{equation}\label{eq:24}
  {{h}_\text{th}}\sim {1.89}{{D}_{0}}{{\We}^{{-3}/{4}}}{\Oh}.
\end{equation}
The above scaling in Eqs.\ (\ref{eq:22}) and (\ref{eq:24}) cannot be compared directly with experimental data due to the difficulties to measure the film thickness from the experimental images. They are used in the further analysis of the singular jet, which will be validated against experimental data later.

\subsection{Retraction stage}\label{sec:33}
\subsubsection{Collapse of the cavity}\label{sec:331}
\begin{figure}
  \centering
  \includegraphics[width=\columnwidth]{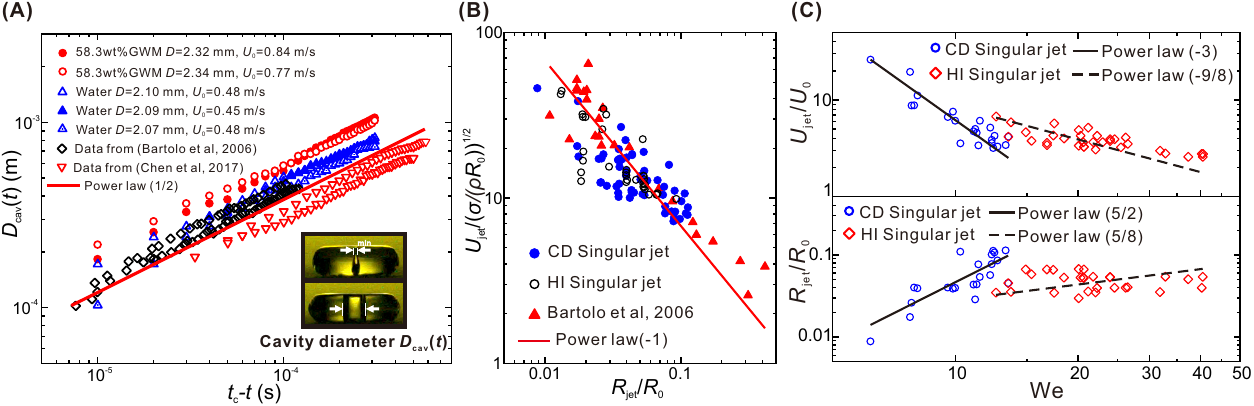}
  \caption{(A) Cavity diameter as a function of time ${{t}_\text{{c}}}-t$, where ${{t}_\text{c}}$ is the collapse time. The circles and the triangles are experimental data of GMW (58.3wt\%) and water respectively. Rhombus and inverted triangles are data reported by Bartolo et al \cite{Bartolo2006} and Chen et al  \cite{Chen2017}, respectively. The solid line is ${{D}_\text{cav}(t)}={0.04}{{\left( {{t}_\text{c}}-t \right)}^{{1}/{2}}}$. (B) The jet velocity as a function of the dimensionless jet radius. Data reported in the literature \cite{Bartolo2006} is also included for comparison. The solid line is ${{U}_\text{jet}}{{[{\sigma} /({\rho} {{R}_{0}})]}^{-1/2}}={0.678}{{\left( {{R}_\text{jet}}/{{R}_{0}} \right)}^{-1}}$.
  (C) Dimensionless jet velocity ${{{U}_{\text{jet}}}}/{{{U}_{0}}}$ as a function of Weber number. The black solid line is ${{{U}_\text{jet}}}/{{{U}_{0}}}={6187}{{\We}^{-3}}$, and the black dash line is ${{{U}_\text{jet}}}/{{{U}_{0}}}={113.6}{{\We}^{-{9}/{8}}}$, in which the coefficients are fitted from experimental data; Dimensionless jet radius ${{{R}_\text{jet}}}/{{{R}_{0}}}$ as a function of Weber number. The black solid line is ${{{R}_\text{jet}}}/{{{R}_{0}}}={1.47}\times {{10}^{-4}}{{\We}^{{5}/{2}}}$, and the black dash line is ${{{R}_\text{jet}}}/{{{R}_{0}}}={6.68}\times {{10}^{-3}}{{\We}^{{5}/{8}}}$, in which the coefficients are fitted from experimental data.}
  \label{fig:04}
\end{figure}

After the formation of the cavity, the cavity collapses quickly. The collapse of the cavity can be described by the Rayleigh-Plesset equation \cite{Bartolo2006, Chen2017} for cylindrical free surface flows
\begin{equation}\label{eq:25}
  {\frac{P(r)-{{P}_{0}}}{\rho }}=\left( {{{\ddot{R}}}_\text{cav}(t)}{{R}_\text{cav}(t)}+\dot{R}_\text{cav}^{2}(t) \right)\ln \left( \frac{{{R}_\text{cav}}(t)}{r} \right)+\frac{1}{2}\dot{R}_\text{cav}^{2}(t)-\frac{{\sigma} +{2}{\mu } {{{\dot{R}}}_\text{cav}}(t)}{\rho {{R}_\text{cav}(t)}},
\end{equation}
where $P(r)$ is the pressure at the radial position of ${r}$, ${{P}_{0}}$ is the pressure in the cavity, and ${{R}_\text{cav}(t)}$ is the radius of the cavity, and the dot indicates the derivative with respect to time. For low-viscosity fluids, the cavity collapses are dominated by inertia force, and the logarithm term in Eq.\ (\ref{eq:25}) can be neglected. Therefore, the relationship between cavity diameter ${{D}_\text{cav}(t)}$ and the collapse time ${{t}_\text{{c}}}$ is ${{D}_\text{cav}(t)}\sim {2}{{\left( {\sigma {{R}_{0}}}/{\rho } \right)}^{{1}/{4}}}{{\left( {{t}_\text{c}}-t \right)}^{{1}/{2}}}$ (see Refs.  \cite{Bartolo2006, Chen2017} for the details). However, for high-viscosity fluids, the cavity collapses are dominated by viscous force, and the last term in Eq.\ (\ref{eq:25}) diverges. In this condition, the relationship between cavity diameter and the collapse time is ${{D}_\text{cav}(t)}\sim \left( {\sigma }/{\mu } \right)\left( {{t}_\text{c}}-t \right)$ (see Refs.  \cite{Bartolo2006, Chen2017} for the details). In our experiment, the Ohsorge number and the Weber numbers are respectively in the ranges of ${\Oh}={2.03}\times {{10}^{-3}}\sim {2.01}\times {{10}^{-2}}$ and ${\We}={3.1}\sim {53.3}$. Hence, we have ${2}{{\left( {\sigma {{R}_{0}}}/{\rho } \right)}^{{1}/{4}}}{{\left( {{t}_\text{c}}-t \right)}^{{1}/{2}}}\ll \left( {\sigma }/{\mu } \right)\left( {{t}_\text{c}}-t \right)$ in this study, meaning that the collapse of the cavity in this study is determined by the inertia rather than the viscosity force.
By comparing the scaling ${{D}_\text{cav}(t)}\sim {2}{{\left( {\sigma {{R}_{0}}}/{\rho } \right)}^{{1}/{4}}}{{\left( {{t}_\text{c}}-t \right)}^{{1}/{2}}}$ with our experimental data of cavity diameters, we can find that the scaling agrees well with our experimental data for both types of singular jets, as shown in Figure \ref{fig:04}(A). The experimental data of the CD singular jets reported by Bartolo et al.\ in Ref.\  \cite{Bartolo2006} and by Chen et al.\ in Ref.\  \cite{Chen2017} are also included in Figure \ref{fig:04}(A), which also shows good agreement. This confirms that the collapse of the cavity is determined by the inertia and the surface tension force.

The characteristic cavity collapse velocity can be obtained by taking derivatives of the maximum cavity radius with respect to time \cite{Bartolo2006}, which leads to
\begin{equation}\label{eq:26}
  {{U}_\text{cav,c}}\sim {{\dot{R}}_\text{cav}({{t}_\text{max}})}\sim \frac{1}{2}{{\left( {\sigma {{R}_{0}}}/{\rho } \right)}^{{1}/{4}}}{{\left( {{t}_\text{c}}-t \right)_\text{max}^{{-1}/{2}}} }\sim {{\left( {\sigma {{R}_{0}}}/{\rho } \right)}^{{1}/{2}}}{{D}_\text{cav,m}^{-1}}.
\end{equation}
The inverse proportionality with respect to ${{D}_\text{cav,m}}$ in Eq.\ (\ref{eq:26}) means that cavity retraction is an accelerated process. As the cavity diameter decreases, the corresponding cavity collapse velocity increases. For CD and HI singular jets, the maximum cavity diameters have been obtained in Eqs.\ (\ref{eq:15}) and (\ref{eq:20}). Therefore, for CD singular jets, substituting Eq.\ (\ref{eq:15}) into Eq.\ (\ref{eq:26}), the characteristic cavity collapse velocity can be obtained as
\begin{equation}\label{eq:27}
  {{U}_\text{cav,c}}\sim {6.33}{{\left( \frac{\sigma }{{\rho}{{R}_{0}} } \right)}^{{1}/{2}}}{{\We}^{-1}}.
\end{equation}
For HI singular jets, substituting Eq.\ (\ref{eq:20}) into Eq.\ (\ref{eq:26}), the characteristic cavity collapse velocity can be obtained as
\begin{equation}\label{eq:28}
  {{U}_\text{cav,c}}\sim {0.94}{{\left( \frac{\sigma }{{\rho}{{R}_{0}} } \right)}^{{1}/{2}}}{{\We}^{-{1}/{4}}}.
\end{equation}
According to Eqs.\ (\ref{eq:27}) and (\ref{eq:28}), the characteristic cavity collapse velocity is determined by the cavity radius for CD singular jet and HI singular jet, which verifies that the cavity formation in spreading stage is important to the cavity collapse in retraction stage. They are used in the further analysis of the singular jets, which will be validated against experimental data later.

\subsubsection{Singular jet velocity and radius}\label{sec:332}
As the cavity collapses, the thin film is the only route where the liquid flows from the rim to the center. The fluid changes from a state of being difficult to enter the film to a state having a large flow velocity in the film and finally producing a high-speed jet. The process for the liquid flow into the liquid film can be regarded as a shape relaxation process (i.e., a process where the system out of equilibrium returns to the equilibrium state \cite{Roland2008}). We can estimate the shape relaxation time \cite{Moran2003, Singh2022, Barakat2023} for the liquid film by analyzing the forces involved. For the film flow, the shape relaxation depends on the balance of inertial and viscous forces in the film, and the timescale of the inertial-viscous balance is ${{t}_\text{relax}}\sim {{\rho}{{R}_\text{cav,m}^{2}}/{\mu}}$ (i.e., the shape relaxation timescale). The observation time for the film flowing is the cavity collapse time ${{t}_\text{c}-{t}}\sim {{{R}_\text{cav,m}^{2}}/{{\left( {\sigma {{R}_{0}}}/{\rho } \right)}^{{1}/{2}}}}$ (based on Eq.\ (\ref{eq:26})). Therefore, we can define a Deborah number as the ratio of the shape relaxation timescale of the film and the observation timescale ${\De}={{t}_\text{relax}}/({{t}_\text{c}-{t}})$. The Deborah number is often used for viscoelastic fluids, but it can also quantify the relaxation of other processes. By substituting their expressions into the definition of ${\De}$, we have
\begin{equation}\label{eq:29}
  {\De}=\frac{{\rho }{{\left( {\sigma {{R}_{0}}}/{\rho } \right)}^{{1}/{2}}}}{\mu}= ({\frac{1}{2}})^{{1}/{2}}{\Oh}^{-1}.
\end{equation}
which is indeed inversely proportional to the Ohnesorge number. This makes sense because the Ohnesorge number can be regarded as a ratio between the capillary timescale and the viscous-inertial timescale. In this study, the Oh number is in the range of ${2.03}\times {{10}^{-3}}$ to ${4.14}\times {{10}^{-2}}$, and the Deborah number is in the range of ${348}$ to ${8.41}\times {{10}^{3}}$ (much higher than 1). Therefore, during the formation of singular jets, in the short observation timescale, the interface of the liquid film does not have time to evolve to achieve an equilibrium between the viscous and inertia. The thickness of the liquid film cannot change greatly under the shape relaxation in the cavity collapse stage. Then, the surface singularity induces the surface tension force suddenly increase which can overcome the large viscous force of the flow in the thin film, eventually forming a high-speed jet.

Even though the flow of the liquid in the liquid film has some similarities with the flow in capillaries \cite{Sorbie1971, Sorbie1995, Ridgway2001} (e.g., the narrow space in the perpendicular direction and the strong viscous force), there are remarkable differences in the formation mechanisms and the role of inertial force. Capillary flow is controlled by surface tension force and viscous force, which is difficult to generate high inertial flow. While for free surface flow in the liquid film in this study, inertial forces play an important role. Therefore, we use mass and momentum conservation to approximately describe the process of liquid flowing into a thin film and the formation of singular jets.

The velocity and the radius of the singular jets are two important parameters to quantify the jet characteristics. The data of the two types of singular jets are used as shown in Figure \ref{fig:04}(B). Regarding the relationship between the velocity ${{U}_\text{jet}}$ and radius ${{R}_\text{jet}}$ of the singular jet, the model proposed by Bartolo et al.\ \cite{Bartolo2006} is adopted. By ignoring the capillary effect and the viscosity, the model assumes: (1) the central film is thin and maintains mass conservation, and (2) the film edge velocity is determined by the cavity collapse velocity. Using the conservation of mass and momentum in the central film
\begin{equation}\label{eq:30}
  \left( {2}{\pi} {{R}_\text{cav,m}}{H} \right){{U}_\text{cav,c}}=\left( {\pi} {{R}_\text{cav,m}^{2}} \right){{U}_\text{jet}},
\end{equation}
\begin{equation}\label{eq:31}
 {\frac{1}{2}}{\rho} { {{U}_\text{cav,c}^{3}}}\left( {2}{\pi} {{R}_\text{cav,m}}{H} \right)={\frac{1}{2}}{\rho} {{U}_\text{jet}^{3}}\left( {\pi} {{R}_\text{jet}^{2}}  \right),
\end{equation}
and combined with Eq.\ (\ref{eq:26}). Substituting Eq.\ (\ref{eq:30}) into Eq.\ (\ref{eq:31}), the relationship between the velocity and the radius of the singular jet can be obtained \cite{Bartolo2006, Chen2017}
\begin{equation}\label{eq:32}
  \frac{{{U}_\text{jet}}}{{{\left( \frac{\sigma }{{\rho} {{R}_{0}}} \right)}^{{1}/{2}}}}\sim {\frac{1}{2}}{{\left( \frac{{{R}_\text{jet}}}{{{R}_{0}}} \right)}^{-1}}.
\end{equation}

Then, the model in Eq.\ (\ref{eq:32}) is compared with our experimental data and the results by Bartolo et al.\ \cite{Bartolo2006}, as shown in Figure \ref{fig:08}. From the comparison, we can see that the model is applicable to both types of singular jets, further validating the assumptions of the thin film and the mass conservation. The thicknesses of the liquid film, even though different for the two types of singular jets, are both much smaller than the droplet thickness during the formation of the singular jets in the whole retraction stage. Therefore, the mass flow rate of the cavity collapse into the liquid film is determined by the droplet thickness.

For CD and HI singular jets, the maximum cavity diameter ${{D}_\text{cav,m}}$ (i.e., Eqs.\ (\ref{eq:15}) and (\ref{eq:20})), the characteristic cavity collapse velocity ${{U}_\text{cav,c}}$ (i.e., Eqs.\ (\ref{eq:27}) and (\ref{eq:28})) and the droplet thickness ${H}$ (i.e., Eqs.\ (\ref{eq:12}) and (\ref{eq:18})) have been derivation, which are the characteristic values in retraction stage. Substituting them into the mass and momentum conservation equations (Eqs.\ (\ref{eq:30}) and (\ref{eq:31})), the characteristic velocities and radiuses of the CD and HI singular jets can be derived respectively. For CD singular jet, the characteristic jet velocity and radius are as follows
\begin{equation}\label{eq:33}
  {{{U}_\text{jet}}}/{{{U}_{0}}}\sim {516.72}{{\We}^{-3}},
\end{equation}
\begin{equation}\label{eq:34}
  {{{R}_\text{jet}}}/{{{R}_{0}}}\sim {1.37}\times {{10}^{-3}}{{\We}^{{5}/{2}}}.
\end{equation}
For HI singular jet, the characteristic jet velocity and radius are as follows
\begin{equation}\label{eq:35}
  {{{U}_\text{jet}}}/{{{U}_{0}}}\sim {4.41}{{\We}^{{-9}/{8}}},
\end{equation}
\begin{equation}\label{eq:36}
  {{{R}_\text{jet}}}/{{{R}_{0}}}\sim {0.16}{{\We}^{{5}/{8}}}.
\end{equation}
The experimental data are fitted by the model in Eqs.\ (\ref{eq:33}) -- (\ref{eq:36}), and show reasonable agreement, as shown in Figure \ref{fig:04}(C). Before singular jet formation, the central liquid film is very thin and the time of the cavity collapse is very small (${\textless}$ ${{1} \text{ms}}$), as shown in Figure \ref{fig:02}. Therefore, the liquid from the rim is difficult to flow into the liquid film (${\De}\gg {1}$) in such a short time. The liquid film is squeezed by the rim forming a jet. As the Weber number increases, the cavity radius increases, the corresponding characteristic cavity collapse velocity decreases, the singular jet velocity decreases, and the singular jet radius increases. For the CD singular jet, the capillary wave propagation induces the capillary deformation of the droplet surface at a low Weber number. The intense deformation makes a spire formation, which moves rapidly downward under the capillary forces (i.e., including the influence of inertial force and surface tension) inducing a very thin central liquid film. However, with the increase of the We number, the spire gradually disappears. The inertial force of the droplet determines the formation of the central liquid film. Therefore, the key force of the central liquid film formation changes from the inertial force and surface tension to the inertial force. In contrast, for HI singular jet, the liquid film only depends on the inertial force. Therefore, the change of liquid film thickness and length for CD singular jets will be faster than that for HI singular jets, which leads to a faster change of the jet velocity and radius of CD singular jets than that of HI singular jets.

\subsection{Regime map of singular jets}\label{sec:34}
\subsubsection{Summary of the regimes}\label{sec:341}

\begin{figure}
  \centering
  \includegraphics[width=0.5\columnwidth]{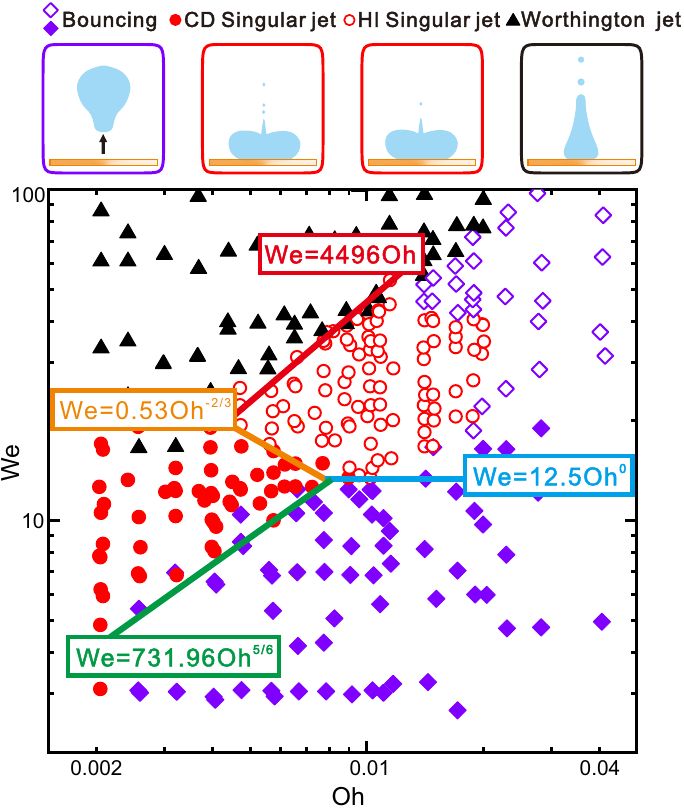}
  \caption{Regime map for singular jets during droplet impact. There are four phenomena, including two types of bouncing, the CD singular jet, the HI singular jet, and the Worthington jet. The green, blue, orange, and red lines respectively represent the critical condition of “bouncing” transforms into “CD singular jet”; “bouncing” transforms into “HI singular jet”; “CD singular jet” transforms into “HI singular jet”; “HI singular jet” transform into “Worthington jet”.}\label{fig:05}
\end{figure}

To see when singular jets occur during the impact process, we performed a large number of experiments by varying the Weber number and the Ohnesorge number of the droplet. As shown in Figure \ref{fig:05}, there are mainly four types of outcomes after the impact, including CD singular jets, HI singular jets, Worthington jets, and two types of bouncing. Singular jets occur only in a limited region of the parameter space, sandwiched between the regions of bouncing and Worthington jets. From the analysis of the droplet spreading and the retraction stages in Sections \ref{sec:33} and \ref{sec:34}, we can know that the formation of the liquid film and the cavity is important for the occurrence of the singular jets. Hence, we will focus on them in the analysis of the transition in the subsequent sections.

\begin{figure}
  \centering
  \includegraphics[width=0.9\columnwidth]{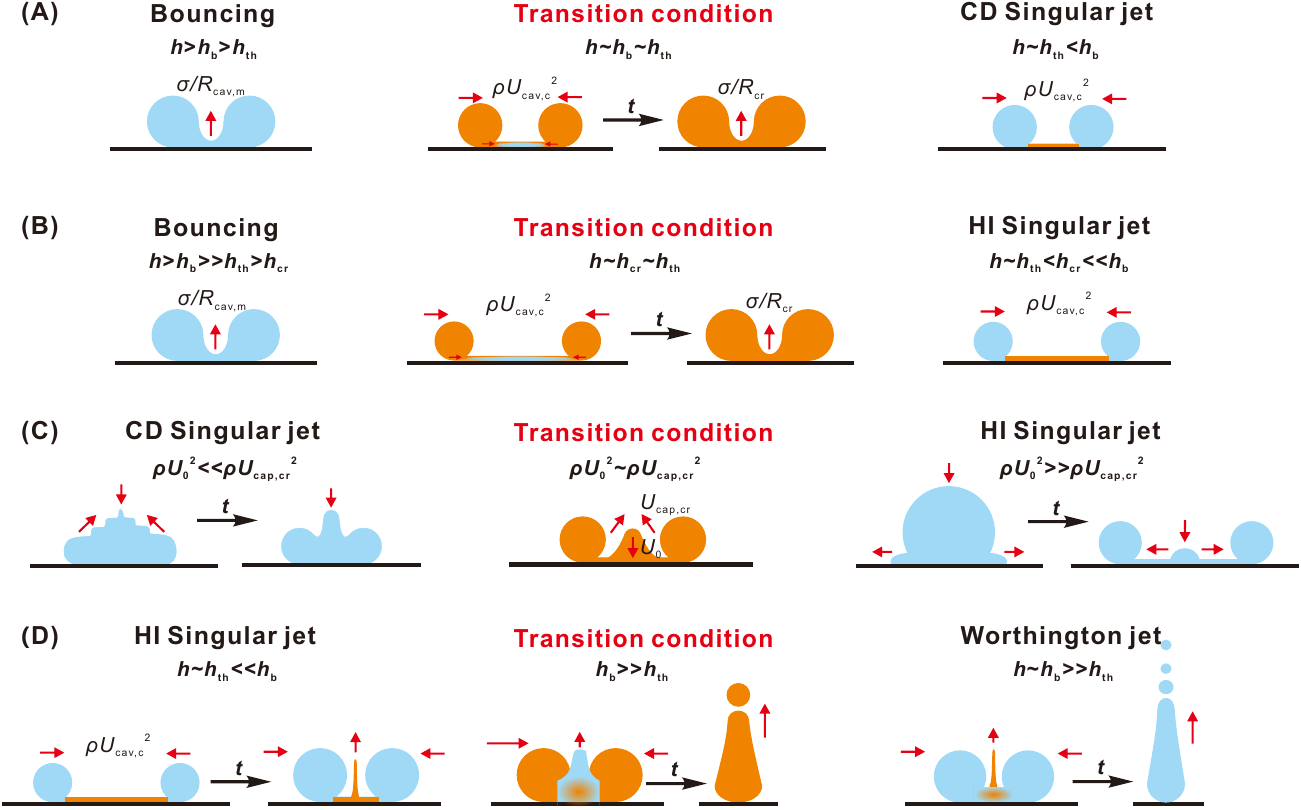}
  \caption{Schematic illustration of the transition conditions. (A) Between bouncing and CD singular jet. (B) Between bouncing and HI singular jet. (C) Between CD and HI singular jets. (D) Between HI singular jet and Worthington jet.}\label{fig:06}
\end{figure}
\subsubsection{Transition from bouncing to CD singular jet}\label{sec:342}
For the transition from bouncing to CD singular jet, the Weber number gradually increases with the Ohnesorge number (see the green line in Figure \ref{fig:05}). At a low Weber number and a high Ohnesorge number, the liquid film is not very thin (see the illustration in Figure \ref{fig:06}(A, Left)) because the droplet does not have sufficient inertia (because of low ${\We}$) to induce strong deformation but the viscous resistance is large (because of large ${\Oh}$). Therefore, the film thickness ${h}$ is not much smaller than maximum cavity radius ${{R}_{\text{cav,m}}}$ (obtained in Eq.\ (\ref{eq:15})), but is much larger than the viscous boundary layer thickness ${{h}_{\text{b}}}$, which follows
\begin{equation}\label{eq:37}
  {{h}_{\text{b}}}\sim \sqrt{{\nu} {{\tau }_\text{i}}}\sim {{D}_{0}}{{\We}^{{-1}/{4}}}{{\Oh}^{{1}/{2}}},
\end{equation}
where ${{\tau }_\text{i}}={{D}_{0}}/{{{U}_{0}}}$. Therefore, the fluid in the toroidal rim can easily flow into the film in the retraction stage. As a consequence, the droplet will be bouncing on the surface. In contrast, at a higher Weber number and a lower Ohnesorge number, the droplet inertia is large, but the viscous effect is low. Therefore, the liquid film becomes very thin (see the illustration in Figure \ref{fig:06}(A, right)). The film thickness ${h}$ is close to the theoretical film thickness ${{h}_{ \text{th}}}$ for the CD singular jet (i.e., Eq.\ (\ref{eq:22})). Therefore, the fluid in the toroidal rim cannot flow into the liquid film. As a consequence, the toroidal rim retracts horizontally under the capillary force while keeping the liquid film thin. Then the formation and the collapse of the cavity induce a CD singular jet.

From the above analysis, we can see that the thickness of the liquid film before the retraction of the toroidal rim is crucial for the transition from bouncing to CD singular jet. If the liquid film is thin, the liquid in the toroidal rim cannot flow into the film, and a CD singular jet occurs; while if the liquid film is thick, the liquid in the toroidal rim can easily enter the film and induce droplet bouncing. Since whether the liquid in the toroidal rim can enter the film depends on the thickness of the viscous boundary layer ${{h}_{\text{b}}}$ and the theoretical film thickness ${{h}_{\text{th}}}$, the critical condition should be that the theoretical film thickness ${{h}_{\text{th}}}$ (i.e., Eq.\ (\ref{eq:22})) is comparable with the viscous boundary layer thickness ${{h}_{\text{b}}}$ (i.e., Eq.\ (\ref{eq:37}))
\begin{equation}\label{eq:38}
  {{h}_\text{th}}\sim {{h}_{\text{b}}}.
\end{equation}
By substituting Eqs.\ (\ref{eq:22}) and (\ref{eq:37}) into Eq.\ (\ref{eq:38}), we can get
\begin{equation}\label{eq:39}
  {33.42}{{D}_{0}}{{\We}^{{-5}/{4}}}{{\Oh}^{{4}/{3}}}\sim {{D}_{0}}{{\We}^{{-1}/{4}}}{{\Oh}^{{1}/{2}}}.
\end{equation}
After simplification, we have
\begin{equation}\label{eq:40}
  {\We}\sim {33.42}{{\Oh}^{{5}/{6}}}.
\end{equation}
As shown in Figure \ref{fig:05}, after the fitting of the prefactor, ${\We}={731.96}{{ \Oh}^{{5}/{6}}}$, the model agrees well with the experimental data.

\subsubsection{Transition from bouncing to HI singular jet}\label{sec:343}
For the transition from bouncing to HI singular jet, the Weber number does not vary with the Ohnesorge number (see the blue line in Figure \ref{fig:05}). At a low Weber number and a high Ohnesorge number, the liquid film is not very thin because the droplet does not have sufficient inertia (because of low ${\We}$) to induce inertia collapse of the toroidal rim but the viscous resistance is large (because of large ${\Oh}$), as illustrated in Figure \ref{fig:06}(B). Therefore, the film thickness ${h}$ is not much smaller than the maximum cavity radius ${{R}_\text{cav,m}}$ (obtained in Eq.\ (\ref{eq:20})), but is much larger than the viscous boundary layer thickness ${{h}_{\text{b}}}$ (i.e., Eq.\ (\ref{eq:37})). Therefore, the fluid in the toroidal rim can easily flow into the film in the retraction stage. As a consequence, the droplet will bounce on the super-hydrophobic surface. In contrast, at a higher Weber number and the same Ohnesorge number, the droplet inertia is large. Therefore, the film becomes thin and the film thickness ${h}$ is close to the theoretical film thickness ${{h}_\text{th}}$ for HI singular jets (i.e., Eq.\ (\ref{eq:24})). However, the large inertia can increase the ability of the toroidal rim liquid to flow into the film. Therefore, the film thickness ${h}$ is smaller than the viscous boundary layer thickness ${{h}_{\text{b}}}$ (i.e., Eq.\ (\ref{eq:37})) and at least close to a critical film thickness ${{h}_\text{cr}}$ (which will be obtained later in Eq.\ (\ref{eq:43})), under which condition, the fluid in the toroidal rim just cannot flow into the film. As a consequence, the toroidal rim retracts horizontally under the capillary force while keeping the film thin. Then the formation of a cavity and the subsequent collapse induces a HI singular jet.

From the above analysis, it can be seen that the thickness of the liquid film before the retraction of the toroidal rim is crucial for the transition from bouncing to HI singular jet. If the liquid film is thin, the liquid in the toroidal rim cannot flow into the film, and the HI singular jet occurs; while if the liquid film is thick, the liquid in the toroidal rim can easily enter the film and induce droplet bounce. Since whether the liquid in the toroidal rim can enter the film depends on the thickness of the critical film thickness ${{h}_\text{cr}}$ and the theoretical film thickness ${{h}_\text{th}}$ (i.e., Eq.\ (\ref{eq:24})), the critical condition should be that the theoretical film thickness ${{h}_\text{th}}$ is comparable with the critical film thickness ${{h}_\text{cr}}$
\begin{equation}\label{eq:41}
  {{h}_\text{th}}\sim {{h}_\text{cr}}.
\end{equation}
The critical film thickness ${{h}_\text{cr}}$ can be obtained by balancing the inertial force ${\rho} {{U}_\text{cav,c}^{2}}$ and the viscous force ${{\mu} {{U}_\text{cav,c}}}/{{{h}_\text{cr}}}$ at the interface between the toroidal rim and the film:
\begin{equation}\label{eq:42}
  {\rho} {{U}_\text{cav,c}^{2}}\sim {\mu} {\frac{{{U}_\text{cav,c}}}{{{h}_\text{cr}}}}.
\end{equation}
By substituting ${{U}_\text{cav,c}}$ (i.e., Eq.\ (\ref{eq:28})) into (\ref{eq:42}), we can get
\begin{equation}\label{eq:43}
  {{h}_\text{cr}}\sim {0.75}{{D}_\text{0}}{{\We}^{{1}/{4}}}{\Oh}.
\end{equation}
By substituting Eqs.\ (\ref{eq:24}) and (\ref{eq:43}) into (\ref{eq:41}) and simplification, we can get
\begin{equation}\label{eq:44}
  {\We}\sim {2.53}{{\Oh}^{0}}.
\end{equation}
As shown in Figure \ref{fig:05}, after the fitting of the prefactor, ${\We}={12.50}{{ \Oh}^{0}}$, the model agrees well with the experimental data.

\begin{figure}
  \centering
  \includegraphics[width=\columnwidth]{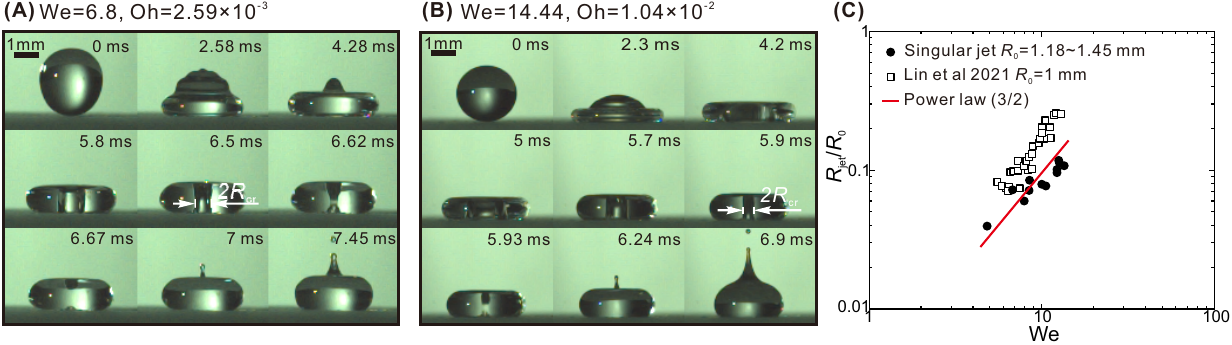}
  \caption{Jet emission in the conditions where the rebound is transformed into a CD or HI singular jet. There is a critical cavity collapse radius ${{R}_\text{cr}}$ in the cavity collapse process. When the cavity shrinks to ${{R}_\text{cr}}$, the central film begins to deform and the singular jet begins to form. (A) CD singular jet, ${\We}={6.8}$ and ${\Oh}={{2.59}\times {{10}^{-3}}}$. (B) HI singular jet, ${\We}={14.44}$ and ${\Oh}={{1.04}\times {{10}^{-2}}}$. (C) Dimensionless radius of singular jets as a function of the Weber number. Circles are the data at the transition condition between bouncing and singular jets. Squares represent data in the literature \cite{Lin2021}. The solid line is ${{{R}_\text{jet}}}/{{{R}_{0}}}={3.0}\times {{10}^{-3}} {{\We}^{{3}/{2}}}$.}\label{fig:07}
\end{figure}
To further verify that the transition from bouncing to CD/HI singular jets depends on the film thickness, we check the radius of the singular jet in the transition condition because the film development in the retraction stage will eventually affect the singular jet formation. When singular jets occur, the variation of the singular jet radius and velocity (i.e., Eq.\ (\ref{eq:32})) satisfies the conservation of mass and energy, i.e., the liquid film is thin enough that the liquid in the rim cannot enter the film. Otherwise, the mass and the kinetic energy of the rim will eventually contribute to the variation of the mass and the energy in the jet. Therefore, when the transition from bouncing to CD/HI singular jets occurs, the singular jet radius should be slightly greater (i.e., with slight mass entering) than that at regular conditions of singular jets (i.e., Eq.\ (\ref{eq:32})). This feature can be seen from the high-speed images in Figure \ref{fig:07}(A and B). In addition, the cavity retraction behavior also reflects the transition condition. At the initial step of the retraction, the cavity follows the inertial collapse regime, like the cavity collapse of singular jets. Then, the bottom of the cavity retracts upward when the droplet retracts to the critical cavity radius ${{R}_\text{cr}}$, which follows the capillary collapse regime like the cavity collapse during the bouncing process. The upward retraction of the cavity resembles the retraction of the cavity after the burst of a bubble at the liquid surface, which produces a jet following a ${3/2}$ power law \cite{Chen2017, Lin2021}. Therefore, the radius of our singular jet should also follow a ${3/2}$ power law:
\begin{equation}\label{eq:45}
  {{R}_\text{jet}}\propto {\sigma }/{\left( {\rho} {{U}_\text{jet}^{2}} \right)}\sim {\sigma }/{\left[ {\rho} {{\left( \frac{\sigma }{{\rho} {{U}_{0}}{{R}_{\max }}} \right)}^{2}} \right]}=\frac{{\rho} {{U}_{0}^{2}}{{R}_{\max }^{2}}}{\sigma }\sim {{R}_{0}} {{\We}^{{3}/{2}}},
\end{equation}
where the relationship of ${{U}_\text{jet}}={\sigma }/{\left( {\rho} {{U}_{0}}{{R}_{\max }} \right)}$ is based on the momentum conservation of jet \cite{Lin2021}. The radii of the singular jets ${{R}_\text{jet}}$ in the transition condition in the experiments are plotted against the Weber number in Figure \ref{fig:07}(C), which can confirm the ${3/2}$ power law in Eq.\ (\ref{eq:45}).

\subsubsection{ransition from CD to HI singular jets}\label{sec:344}
For the transition from CD to HI singular jets, the Weber number gradually decreases with the Ohnesorge number (see the orange line in Figure \ref{fig:05}). As illustrated in Figure \ref{fig:06}(C), at a high Weber number (relatively higher than the bouncing regime) and a low Ohnesorge number, the capillary deformation is significant because the droplet has sufficient inertia (i.e., high ${\We}$) but the viscous resistance is low (i.e., low ${\Oh}$). Because the capillary deformation is induced by the capillary wave, the kinetic energy of the surface capillary wave drives the liquid upward. Therefore, the kinetic energy of the surface capillary wave is much larger than the kinetic energy of the top surface. As a consequence, the top surface of the droplet will deform to a spire, and eventually form a CD singular jet. In contrast, at a higher Weber number and a higher Ohnesorge number, the viscous effect is large, which can dampen the capillary deformation. Therefore, the kinetic energy of the surface capillary wave is much smaller than the kinetic energy of the top surface. As a consequence, the top surface of the droplet will not deform, eventually forming a HI singular jet.

From the above analysis, we can see that the disappearance of capillary deformation is crucial for the transition from CD singular jet to HI singular jet. If the kinetic energy of the surface capillary wave is much larger than the kinetic energy of the top surface, the capillary deformation appears and the CD singular jet occurs. Otherwise, the capillary deformation disappears and the HI singular jet occurs. Therefore, the critical condition should be that the kinetic energy of the top surface ${\rho} {{U}_{0}^{2}}$ is comparable with the kinetic energy of the surface capillary wave ${\rho} {{U}_{\text{cap,cr}}^{2}}$
\begin{equation}\label{eq:46}
  {\rho} {{{U}_{0}^{2}}}\sim {\rho} {{U}_\text{cap,cr}^{2}}.
\end{equation}

The critical capillary wave velocity when the pyramid structure disappears ${{U}_\text{cap,cr}}$ in Eq.\ (\ref{eq:46}) can be obtained from the formation and disappearance of the pyramid structure. For the formation stage of the pyramid structure, based on the balance of inertia and surface tension (i.e., ${\rho} {{U}^{2}}\sim {\sigma }/{\lambda }$), we can obtain the capillary wavelength ${\lambda} \sim {\sigma }/{\left( {\rho} {{{U}}^{2}} \right)}$. In addition, the disappearance of the pyramid structure is due to the decay of a wave whose wavelength is determined by the viscous force. By balancing the inertia and the viscous force of the capillary wave (i.e., ${\rho} {{U}^{2}}\sim {{U}{\ell} {\mu} }/{\left( {{\lambda }^{2}} \right)}$ where ${\ell }$ is the decay distance), we can obtain the decay distance ${\ell} \sim {{\rho} {{\lambda }^{2}}{U}}/{\mu }$. When the decay distance ${\ell }$ is greater than the radius of the droplet ${{{R}_{0}}}$, no capillary wave can be observed on the droplet. Therefore, the critical surface wave velocity \cite{Renardy2003} for the disappearance of the pyramid can be obtained by substituting the capillary wavelength into the decay distance.
\begin{equation}\label{eq:47}
  {{U}_\text{cap,cr}}\sim {{\left( \frac{{{\sigma }^{2}}}{{\rho} {\mu} {{R}_{0}}} \right)}^{{1}/{3}}}.
\end{equation}
By substituting Eq.\ (\ref{eq:47}) into Eq.\ (\ref{eq:46}), we can get
\begin{equation}\label{eq:48}
  {\We}\sim {1.59}{{\Oh}^{{-2}/{3}}}.
\end{equation}
As shown in Figure \ref{fig:05}, after fitting of the prefactor, ${\We}={0.53}{{ \Oh}^{{-2}/{3}}}$, the model agrees well with the experimental data.
\subsubsection{Transition from HI singular jet to Worthington jet}\label{sec:345}
For the transition from HI singular jet to Worthington jet, the Weber number gradually increases with the Ohnesorge number (see the red line in Figure \ref{fig:05}). As illustrated in Figure \ref{fig:06}(D), at a high Weber number and a high Ohnesorge number, the liquid film is very thin because the droplet has sufficient inertia (because of high ${\We}$) and viscous resistance (because of large ${\Oh}$). Therefore, the fluid in the toroidal rim cannot flow into the film in the retraction stage. The film thickness ${h}$ is close to the theoretical film thickness ${{h}_\text{th}}$ for the HI singular jet (i.e., Eq.\ (\ref{eq:24})). As a consequence, an HI singular jet will form. In contrast, at a higher Weber number and a lower Ohnesorge number, the droplet inertia is large, but the viscous effect is low. The large inertia further increases the droplet spreading radius ${{{R}_{\max }}}$ (i.e., Eq.\ (\ref{eq:01})), and the maximum cavity radius ${{{R}_\text{cav,m}}}$ (i.e., Eq.\ (\ref{eq:20})) is close to the droplet spreading radius ${{{R}_{\max }}}$, which is much larger than the initial radius of the droplet ${{{R}_{0}}}$. During the flattening of the upper surface of the droplet by the large inertia, after the upper surface has been completely flattened, the large inertia makes the droplet continue spreading in the horizontal direction. The film thickness ${h}$ is determined by the viscous boundary layer thickness ${{h}_{\text{b}}}$ (i.e., Eq.\ (\ref{eq:37})) on the free surface. Therefore, the film is thicker than that in HI singular jet conditions. Such variation of the film thickness was also founded by Zhang et al.\ \cite{Zhang2022} when the Weber number increases from the singular jet regime to the Worthington jet regime. Therefore, the fluid in the toroidal rim can flow into the film, and the toroidal rim retracts horizontally, forming a thick jet (i.e., the Worthington jet).

From the above analysis, it can be seen that the thickness of the liquid film before the retraction of the toroidal rim is crucial for the transition from the HI singular jet to the Worthington jet. If the liquid film is thin, the liquid in the toroidal rim cannot flow into the film, and the HI singular jet occurs; while if the liquid film is thick, the liquid in the toroidal rim can enter the film and induce the Worthington jet. Whether the liquid in the toroidal rim can enter the film and induce a thick jet depends on the theoretical film thickness ${{h}_\text{th}}$ (i.e., Eq.\ (\ref{eq:24})) and the thickness of the viscous boundary layer ${{h}_{\text{b}}}$ (i.e., Eq.\ (\ref{eq:37})). According to the experimental data in Figure \ref{fig:05}, we know that the viscous boundary layer thickness ${{h}_{\text{b}}}$ (i.e., Eq.\ (\ref{eq:37})) is much larger than the theoretical film thickness ${{h}_\text{th}}$ (i.e., Eq.\ (\ref{eq:24})). Therefore, the critical condition should be
\begin{equation}\label{eq:49}
  {{h}_{\text{b}}}\gg {{h}_\text{th}}.
\end{equation}
By substituting Eqs.\ (\ref{eq:24}), (\ref{eq:37}) into (\ref{eq:49}), we can get
\begin{equation}\label{eq:50}
  {\We}\gg {\Oh}.
\end{equation}
Hence, we obtain a very large prefactor in the fitting of the experiment data, ${\We}={4496} {\Oh}$, as shown in Figure \ref{fig:05}.

\subsubsection{Transition from HI singular jet to bouncing}\label{sec:346}
\begin{figure}
  \centering
  \includegraphics[width=0.7\columnwidth]{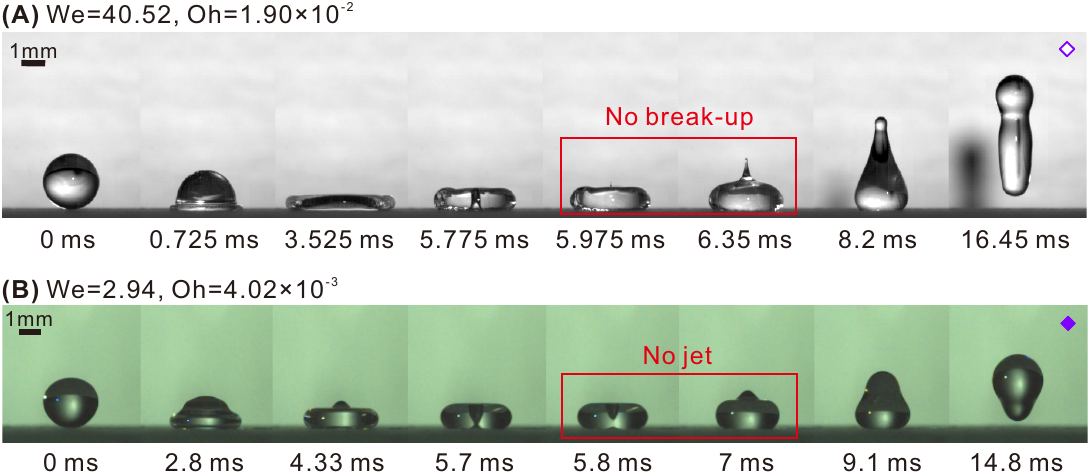}
  \caption{Two types of bouncing processes at different impact conditions of ${\We}$ and ${\Oh}$. (A) The liquid column is not strong enough to break up because of the higher fluid viscosity, corresponding to the purple open rhombus in Figure \ref{fig:05}. (B) The retraction of the droplet’s upper surface is not strong enough to induce an obvious liquid column, corresponding to the purple filled rhombus in Figure \ref{fig:05}. Video clips of these processes are available as Supplementary Material as Movies 3 and 4.}\label{fig:08}
\end{figure}

As shown in Figure \ref{fig:05}, as the ${\Oh}$ number increases, the impact will transit from HI singular jets to bouncing, as shown in Figure \ref{fig:08}(A). For the transition from the HI singular jet to the bouncing, a liquid column is first formed at the top surface of the droplet during retraction. However, With the increases of the ${\Oh}$ number, the viscous force also increases. Therefore, the liquid column does not break up (see ${5.975}$ -- ${6.35}$ ${\text{ms}}$ in Figure \ref{fig:08}(A), and bouncing finally occurs. The bouncing process is different from the bouncing process at a low Weber number (see Figure \ref{fig:08}(B)), during which the droplet’s upper surface only retracts but is not strong enough to induce an obvious liquid column.

\definecolor{color1}{RGB}{128, 0, 255}

\section{Conclusions}\label{sec:4}
In this study of droplet impact on super-hydrophobic surfaces, we find a new type of singular jet (i.e., HI singular jets), besides the previously studied CD singular jets \cite{Bartolo2006, Chen2017, Lin2021, Lin2018}. They have major differences in cavity formation, which retracts to produce singular jets. For the CD singular jet, the cavity formation is due to the propagation of the capillary wave on the surface of the droplet, which induces strong surface vibration at the apex of the droplet; for the HI singular jet, the cavity formation is by the thin central film and the toroidal edge, which has large horizontal inertia during the retraction of the droplet after the maximum spreading. The process of singular jet formation is analyzed, including the spreading and retraction stages of droplet impact. Key parameters in the impact process are identified and analyzed quantitatively such as the droplet spreading diameter, the shape and the speed of the spire, the size and retraction velocity of the cavity, the center film thickness, and the radius and velocity of the jets. Finally, the boundaries for the formation of singular jets in the ${\Oh}$--${\We}$ parameter space are analyzed, and scaling relationships for the transition conditions are obtained. Since the impact of droplets is a fundamental process of fluid mechanics, the results of this study will be useful for us to understand the mechanism in numerous scenarios, such as the generation and dispersion of aerosol by raindrops \cite{2015Aerosol} or respiratory activities \cite{2021respiratory}. In addition, the results of this study are also helpful for controlling droplet behaviors in numerous applications, such as inkjet printing, additive manufacturing, and spray painting. For example, because the singular jet is very thin and very fast, the impact of droplets on super-hydrophobic surfaces can be used as potential methods for needle-free fluid injection \cite{NeedleFree2006} and high-resolution fluid dispensing \cite{Lin2021}. Considering practical applications, the singular jets during droplet impact deserves more studies, such as the effects of non-Newtonian rheologies and additives, e.g., polymers, surfactants, and particles.

\section*{Declaration of Competing Interest}
The authors declare that they have no known competing financial interests or personal relationships that could have appeared to influence the work reported in this paper.

\section*{Acknowledgements}
This work is supported by the National Natural Science Foundation of China (Grant Nos.\ 51676137, 52176083, and 51920105010).

\section*{Appendix A. Supplementary material}
Supplementary material associated with this article can be found in the online version.

\bibliography{singularJet}
\end{document}